\documentclass[epj]{webofc}
\usepackage[utf8]{inputenc}
\usepackage[varg]{txfonts}   % Web of Conferences font
\usepackage{booktabs}
\usepackage{xcolor}
\definecolor{darkred}{rgb}{0.4,0.0,0.0}
\definecolor{darkgreen}{rgb}{0.0,0.4,0.0}
\definecolor{darkblue}{rgb}{0.0,0.0,0.4}
\usepackage[bookmarks,linktocpage,colorlinks,
    linkcolor = darkred,
    urlcolor  = darkblue,
    citecolor = darkgreen]{hyperref}
\usepackage{subfigure}
\wocname{EPJ Web of Conferences}
\woctitle{Lattice2017}
\begin{document}
%%%%%%%%%%%%%%%%%%%%%%%%%%%%%%%%%%%%%%%%%%%%%%%%%%%%%%%%%%%%%%%%%%%%%%%%%%%%%
%
\selectlanguage{english}

\hspace{10cm}
\begin{tabular}{r}
{\tt LPT-Orsay-17-57}\\
{\tt MS-TP-17-18}
\end{tabular}

\vspace{-1cm}
%----------------------------------------------------------------------------
\title{%
On the $D^*_s$ and charmonia leptonic decays
}
%----------------------------------------------------------------------------
\author{%
\firstname{Gabriela} \lastname{Bailas}\inst{1} \and
\firstname{Beno\^\i t} \lastname{Blossier}\inst{2}\fnsep\thanks{Speaker, \email{benoit.blossier@th.u-psud.fr}} \and
\firstname{Jochen}  \lastname{Heitger}\inst{3} \and \firstname{Vincent}  \lastname{Mor\'enas}\inst{1}
\and \firstname{Matthias}  \lastname{Post}\inst{3}
% etc.
}
%----------------------------------------------------------------------------
\institute{%
Laboratoire de Physique de Clermont, Campus Universitaire des C\'ezeaux, 4 Avenue Blaise Pascal, TSA 60026, 63171 Aubi\`ere Cedex, France
\and
Laboratoire de Physique Th\'eorique, CNRS, Univ. Paris-Sud et Universit\'e Paris-Saclay,  B\^atiment 210,  91405 Orsay Cedex, France
\and
Institut~f\"ur~Theoretische~Physik, Universit\"at~M\"unster, Wilhelm-Klemm-Str.~9, 48149~M\"unster, Germany
}
%----------------------------------------------------------------------------
\abstract{%
  Among the different scenarios of New Physics, those with an extended Higgs sector are examined with a lot of attention. Recent experimental observations of several anomalies in flavour physics with respect to expectations of the Standard Model further motivate the effort of phenomenologists. First, informations about the $R_{D_s}$ ratio, a test of lepton flavour universality equivalent to $R_D$, already measured, but with the $s$ quark as spectator, are awaited in coming years to constrain the corner of an extended Higgs sector with charged doublets.
On another side, leptonic widths of pseudoscalar quarkonia are particularly interesting to test an extended Higgs sector with a light CP-odd Higgs boson singlet, through the study of its mixing with quarkonia states. Hadronic parameters entering those processes have to be determined from lattice QCD with enough confidence on the control of systematic errors. We report on the very first step of a long-term program tackled with ${\rm N_f}=2$ Wilson-Clover fermions to put relevant constraints on extensions of the Higgs sector: extraction of decay constants of $D^*_s$, $\eta_c$, $\eta_c(2S)$, $J/\psi$ and $\psi(2S)$ with lattice ensembles provided by the CLS effort, considering 2 lattice spacings and a large range of pion masses to estimate cut-off effects and extrapolate results to the chiral limit.
}
%----------------------------------------------------------------------------
\maketitle
%----------------------------------------------------------------------------
\section{Introduction}\label{intro}

The discovery at LHC of the Higgs boson with a mass of 125 GeV has been a major milestone in the history of Standard Model (SM) tests: the spontaneous breaking of electroweak symmetry generates masses of charged leptons, quarks and weak bosons. A well-known issue with the SM Higgs is that the quartic term in the Higgs Lagrangian induces for the Higgs mass $m_H$ a quadratic divergence in the hard scale of the theory: it is related to the so-called hierarchy problem. Several scenarios beyond the SM are proposed to cure the issue. Minimal extensions of the Higgs sector contain two complex scalar isodoublets $\Phi_{1,2}$ that, after the spontaneous breaking of the electroweak symmetry, lead to 2 charged particles $H^\pm$, 2 CP-even particles $h$ (SM-like Higgs) and $H$ and 1 CP-odd particle $A$. 
In that class of scenarios, quarks are coupled to charged Higgs through a right-handed current and to the CP-odd Higgs through a pseudoscalar current. Phenomenological consequences have recently received a lot of attention. On the one hand, several tests of lepton flavour universality have shown some hints of anomaly with respect to the SM expectations, especially for the ratios $R_{D^{(*)}}\equiv \frac{\Gamma(B\to D^{(*)} \tau \nu_\tau)}{\Gamma(B \to D^{(*)} \ell \nu_\ell)}, \ell=e,\mu$ \cite{Lees:2012xj, Huschle:2015rga, Aaij:2015yra}: semileptonic decays with $\tau$ lepton in final states can have a non-SM contribution from the exchange of a right-handed current that is not helicity suppressed. Changing the spectator quark of the $b\to c$ flavour transition, it is worth investigating ratios $R_{D^{(*)}_s}$, for instance at Belle-2, assuming on the theory side a very good control on hadronic properties of $B_s$, $D_s$ and $D^*_s$ mesons.
On the other hand,  the leptonic decay of pseudoscalar quarkonia, highly suppressed in the SM because it occurs \emph{via} quantum loops, can be reinforced by a new tree-level contribution mediated by a light CP-odd Higgs boson \cite{Fullana:2007uq}: any enhanced observation with respect to the SM expectation would be a clear signal of New Physics. Obviously the hadronic inputs to constrain the CP-odd Higgs coupling to heavy quarks are the decay constant of pseudoscalar quarkonia.

%----------------------------------------------------------------------------
\section{Lattice analysis}\label{sec-1}

\begin{figure}[t]
\begin{center}
\begin{tabular}{cc}
\includegraphics[width=4.5cm, clip]{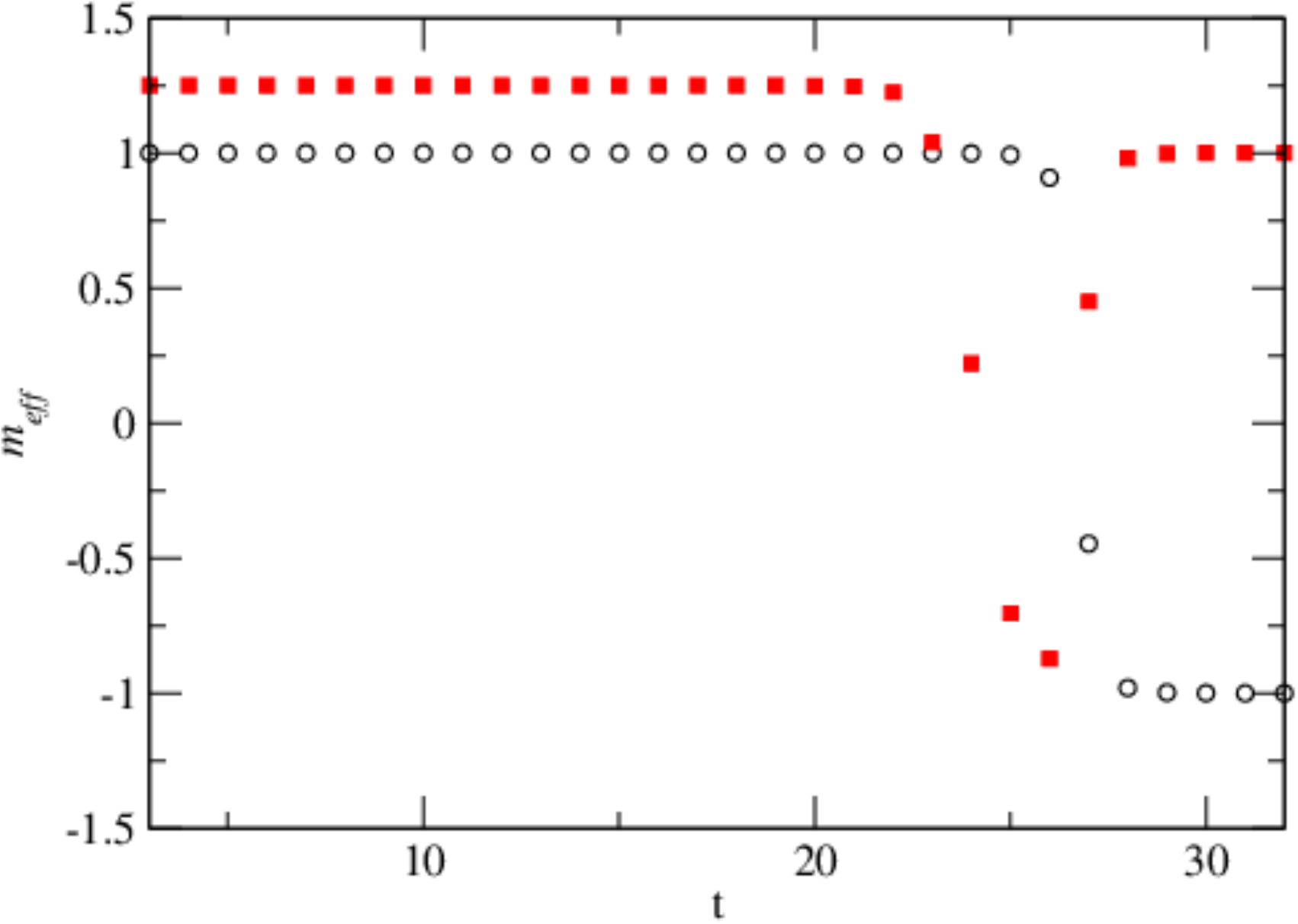}
&	
\includegraphics[width=4.5cm, clip]{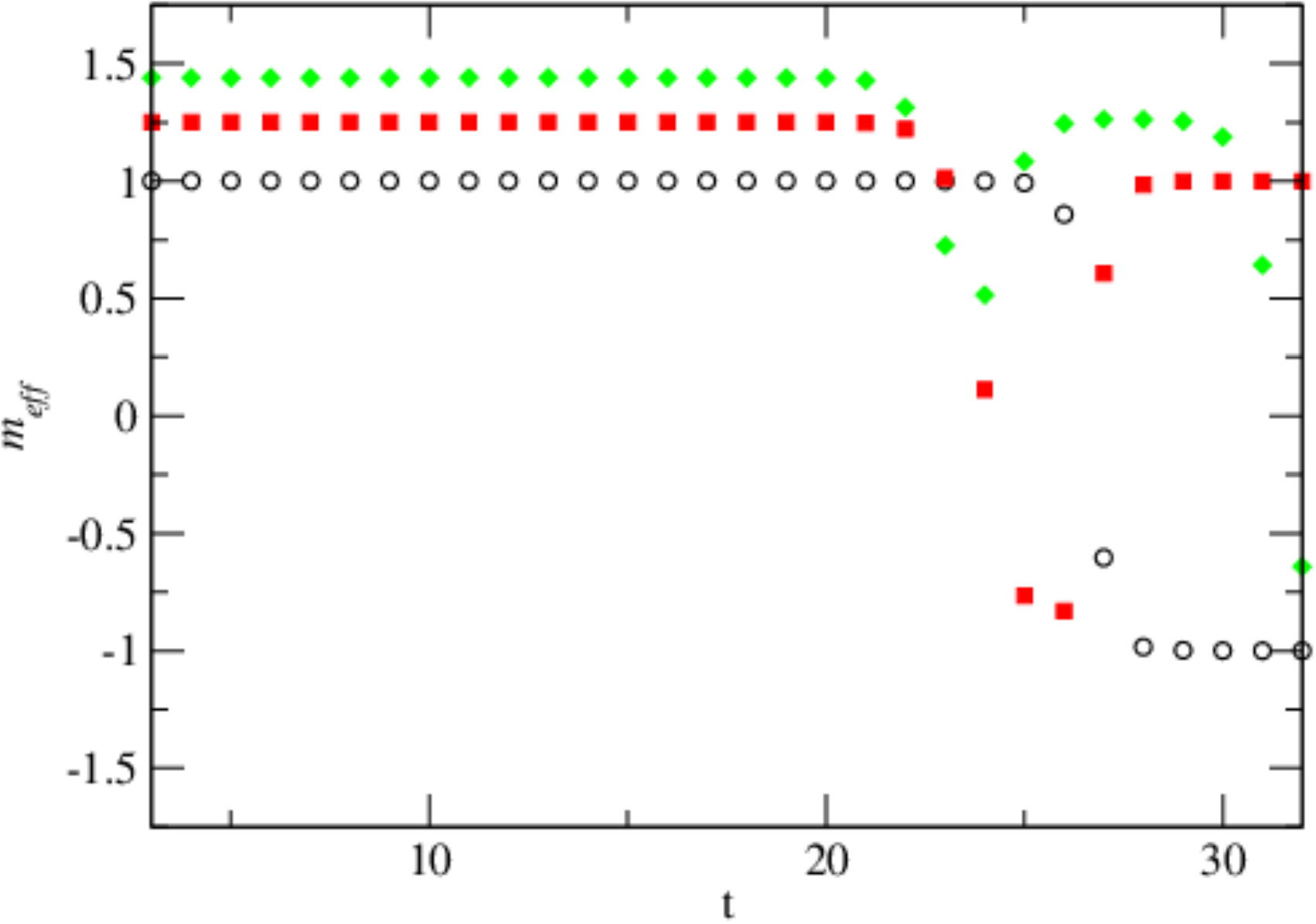}\\
\end{tabular}
\end{center}
	\caption{Effective masses, in lattice units, obtained from the $2\times 2$ subsystem (left panel) and the $3 \times 3$ subsystem (right panel) of our toy model, with $T=64$ and $t_0=3$.}
\label{fig:toymodel}
\end{figure}

Our work has been performed using a subset of the CLS ensembles with ${\rm N_f}=2$ ${\cal O}(a)$ improved Wilson-Clover fermions, whose parameters are collected in Table~\ref{tabsim}. Two lattice spacings $a_{\beta=5.5}=0.04831(38)$ fm and $a_{\beta=5.3}=0.06531(60)$ fm, resulting from a fit in the chiral sector \cite{Lottini:2013rfa}, are considered; we have taken simulations with pion masses in the range $[190\,, 440]$~MeV. The charm quark mass has been tuned after a linear interpolation of $m^2_{D_s}$ in $1/\kappa_c$ at its physical value \cite{Heitger:2013oaa}, after having fixed the strange quark mass \cite{Fritzsch:2012wq}. The statistical error is estimated from the jackknife procedure: 2 successive measurements are sufficiently separated in trajectories along the Monte-Carlo history to neglect autocorrelation effects. We have computed quark propagators in two-point correlation functions using stochastic sources that are different from zero in a timeslice that changes randomly for each measurement; we have applied spin dilution and the one-end trick to reduce the stochastic noise \cite{Foster:1998vw, McNeile:2006bz}.
\begin{table}[t]
\caption{Parameters of the simulations: bare coupling $\beta = 6/g_0^2$, lattice resolution, hopping parameter $\kappa$, lattice spacing $a$ in physical units, pion mass, number of gauge configurations and bare charm quark masses.}
\label{tabsim}
\begin{center}
\begin{tabular}{lcc@{\hskip 02em}c@{\hskip 02em}c@{\hskip 01em}c@{\hskip 01em}c@{\hskip 01em}c@{\hskip 01em}c}
\hline
	\toprule
	id	&	$\quad\beta\quad$	&	$(L/a)^3\times (T/a)$ 		&	$\kappa_{\rm sea}$		&	$a~(\rm fm)$	&	$m_{\pi}~({\rm MeV})$	& $Lm_{\pi}$ 	&$\kappa_s$&$\kappa_c$\\
\hline
	\midrule
	E5	&	5.3		&	$32^3\times64$	& 	$0.13625$	& 	0.065	  	& 	$440$	&4.7	&$0.135777$&$0.12724$\\  
	F6	&			& 	$48^3\times96$	&	$0.13635$	& 			& 	$310$	&5	&$0.135741$&$0.12713$\\    
	F7	&			& 	$48^3\times96$	&	$0.13638$	& 			& 	$270$	&4.3	&$0.135730$&$0.12713$\\    
	G8	&			& 	$64^3\times128$	&	$0.13642$	& 			& 	$190$	&4.1	&$0.135705$&$0.12710$\\    
\hline
	\midrule
	N6	&	$5.5$	&	$48^3\times96$	&	$0.13667$	& 	$0.048$  	& 	$340$	&4	& $9.136250$&$0.13026$\\	
	O7	&		&	$64^3\times128$	&	$0.13671$	& 	 	& 	$270$	&4.2	& $0.136243$&$0.13022$ \\ 
	\bottomrule
\hline
\end{tabular} 
\end{center}
%\caption{Parameters of the simulations: bare coupling $\beta = 6/g_0^2$, lattice resolution, hopping parameter $\kappa$, lattice spacing $a$ in physical units, pion mass, number of gauge configurations and bare charm quark masses.}
%\label{tabsim}
\end{table}
Two-point correlation functions under investigation are
$C_{\Gamma \Gamma'}(t)=\frac{1}{V} \sum_{\vec{x},\vec{y}} \langle [\bar{c}\Gamma Q](\vec{y},t) [\bar{Q}\gamma_0 \Gamma' \gamma_0 c](\vec{x},0)\rangle$,
$Q=c$ or $s$, where $V$ is the spatial volume of the lattice, $\langle ... \rangle$ the expectation value over gauge configurations and interpolating fields $\bar{c} \Gamma Q$ are not always local.  As a preparatory step we have examined different possibilities to find the best basis of operators, combining levels of Gaussian smearing, interpolating fields with a covariant derivative $\bar{c} \Gamma \vec{\gamma}\cdot \vec{\nabla} Q$ and operators that are odd under time parity. Solving the Generalized Eigenvalue Problem (GEVP) \cite{Michael:1985ne, Luscher:1990ck} is a key point in our analysis.\\ 
Looking at the literature on lattice studies of charmonia, we have noticed that people tried to mix together the operators $\bar{c} \Gamma c$ and $\bar{c} \gamma_0 \Gamma c$ in a unique GEVP system \cite{Liu:2012ze, Becirevic:2014rda}: according to us, that approach raises questions. To explain our puzzle, we take the example of the interpolating fields $\{P=\bar{q} \gamma_5 q,\;\; A_0=\bar{q} \gamma_0 \gamma_5 q\}$; we have the following asymptotic behaviours: 
\begin{eqnarray}\nonumber
\langle P(t)P(0) \rangle, \langle A_0(t) A_0(0)\rangle &\stackrel{t\to \infty}{\longrightarrow}&{\rm cosh}[m_P(T/2-t)],\\
\nonumber
\langle P(t)A_0(0) \rangle, \langle A_0(t) P(0)\rangle&\stackrel{t\to\infty}{\longrightarrow} &{\rm sinh}[m_P(T/2-t)].
\end{eqnarray}
The matrix of $2\times 2$ correlators of the GEVP is then 
\begin{equation}\nonumber
C(t)=\left[\begin{array}{cc}\langle P(t)P(0) \rangle&\langle A_0(t)P(0) \rangle\\
\langle P(t)A_0(0) \rangle&\langle A_0(t)A_0(0) \rangle \end{array}\right]\quad {\rm GEVP}: 
C(t) v_n(t,t_0)=\lambda_n(t,t_0) C(t_0) v_n(t,t_0).
\end{equation}
In the general case, the spectral decomposition of $C_{ij}(t)$ is 
\begin{equation}\nonumber
C_{ij}(t)=\sum_n Z^i_n Z^{*j}_n [D_{ij}\rho^{(1)}_n(t) + (1-D_{ij}) \rho^{(2)}_n(t)],\; D_{ij}=0\;{\rm or}\;1, 
\end{equation}
with $\rho^{(1),(2)}(t) \sim e^{-m_Pt}$, ${\rm cosh}[m_P(T/2-t)]$, ${\rm \sinh}[m_P(T/2-t)]$. The dual vector $u_n$ to $Z's$ is defined by
$\sum_j Z^{*j}_m u^j_n = \delta _{mn}$. Inserted in the GEVP, it gives
\begin{eqnarray}\nonumber
\sum_j C_{ij}(t)u^j_n &=&\sum_{j,m} Z^i_m Z^{*j}_m u^j_n [D_{ij}\rho^{(1)}_m(t) + (1-D_{ij}) \rho^{(2)}_m(t) ]\\
%\nonumber
%&=&\sum_m \rho^{(2)}_m(t) Z^i_m \sum_j Z^{*j}_m u^j_n + \sum_m (\rho^{(1)}_m(t)-\rho^{(2)}_m(t))Z^i_m \sum_j D_{ij} Z^{*j}_m u^j_n\\
\nonumber
&=&\rho^{(2)}_n(t) Z^i_n + \sum_m (\rho^{(1)}_m(t)-\rho^{(2)}_m(t))Z^i_m \sum_j D_{ij} Z^{*j}_m u^j_n
\end{eqnarray}
If $D_{ij}$ is independent of $i,j$, we can write 
\begin{equation}\nonumber
C(t)u_n = \rho(t) Z_n,\quad \lambda_n(t,t_0)=\frac{\rho_n(t)}{\rho_n(t_0)}.
\end{equation}
Approximating every correlators by sums of exponentials forward in time may face caveats. A toy model with 3 states in the spectrum helps to understand this issue:
\begin{center}
\begin{tabular}{|c|}
\hline
spectrum\\
\hline
1.0\\
1.25\\
1.44\\
\hline
\end{tabular}
\begin{tabular}{c}
Matrix of couplings\\
$\left[ \begin{array}{ccc}
0.6&0.25&0.08\\
0.61&0.27&0.08\\
0.58&0.24&0.08\\
\end{array}
\right]$
\end{tabular}
\begin{tabular}{c}
time behaviour of $C_{ij}$\\
$\left[ \begin{array}{ccc}
\cosh&\sinh&\cosh\\
\sinh&\cosh&\sinh\\
\cosh&\sinh&\cosh\\
\end{array}
\right]$
\end{tabular}
\end{center}
The effective mass got from solving the GEVP reads
$am_{eff, n}=\ln\left(\frac{\lambda_n(t,t_0)}{\lambda_n(t+a,t_0)}\right)$.
In our numerical application, we have chosen $T/a=64$, $t_0/a=3$ and compared $2\times 2$ and $3 \times 3$ subsystems: results can be seen in Fig. \ref{fig:toymodel}.
Our observation is that until $t=T/4$ there is no effect of neglecting the time-backward contribution in the correlation function. So it is certainly safe for the ground state or the first excitation. On another side one might wonder what might happen with a dense spectrum when one extracts the energy of the $3^{\rm rd}$ or a higher excited state.\\
Building a basis of operators with $\{\bar{c} \Gamma c; \bar{c} \Gamma \vec{\gamma} \cdot \vec{\nabla} c\}$ could be beneficial and it was already explored \cite{Dudek:2010wm, Mohler:2012na}. But there are sometimes bad surprises, a good example is the pseudoscalar-pseudoscalar correlator $C(t)=\langle [\bar{c} \gamma^5 \vec{\gamma} \cdot \vec{\nabla} c](t)
 [\bar{c} \gamma^5 \vec{\gamma} \cdot \vec{\nabla} c](0) \rangle$.
Indeed, we have found
that the ``diagonal" contribution $A(t)=\sum_i \langle [\bar{c} \gamma^5 \gamma_i \nabla_i c](t)[\bar{c} \gamma^5 \gamma_i \nabla_i c](0) \rangle$ cancels with the ``off-diagonal" contribution $B(t)=\sum_{i\neq j} \langle [\bar{c} \gamma^5 \gamma_i \nabla_i c](t)[\bar{c} \gamma^5 \gamma_j \nabla_j c](0) \rangle$, resulting in a correlator $C(t)$ very noisy and compatible with zero.\\
Eventually we have considered 4 Gaussian smearing levels for the quark fields $c$ and $s$, including no smearing, to build $4\times 4$ matrix of correlators without any covariant derivative and no operator of the $\pi_2$ or $\rho_2$ kind \cite{Dudek:2010wm}, from which we also extract the ${\cal O}(a)$ improved hadronic quantities we examine. Solving the GEVP for the pseudoscalar-pseudoscalar
and vector-vector matrices of correlators 
\begin{equation*}
C_{PP}(t)v^P_n(t,t_0)=\lambda^P_n(t,t_0) v^P_n(t,t_0)C_{PP}(t_0), \quad 
C_{VV}(t)v^V_n(t,t_0)=\lambda^V_n(t,t_0) v^V_n(t,t_0)C_{VV}(t_0),
\end{equation*} 
we obtain the correlators that will have the largest overlap with the $n^{\rm th}$ excited state as follows:
\begin{eqnarray}\nonumber
\tilde{C}^n_{A_0 P}(t) &=& \sum_i C_{A^L_0 P^{(i)}}(t) v^{P,i}_n(t,t_0),\\
\nonumber
\tilde{C}^n_{P P}(t) &=& \sum_i C_{P^L P^{(i)}}(t) v^{P,i}_n(t,t_0),\\
\nonumber
\tilde{C}'^n_{P P}(t) &=& \sum_{i,j} v^{P,i}_n(t,t_0) C_{P^{(i)} P^{(j)}}(t) v^{P,j}_n(t,t_0),\\
\nonumber
\tilde{C}^n_{V V}(t) &=& \frac{1}{3} \sum_{i,k} C_{V^L_k V^{(i)}_k}(t) v^{V,i}_n(t,t_0),\\
\nonumber
\tilde{C}'^n_{VV}(t) &=& \frac{1}{3} \sum_{i,j,k} v^{V,i}_1(t,t_0) C_{V^{(i)}_k V^{(j)}_k}(t) v^{V,j}_n(t,t_0),\\
\nonumber
\tilde{C}^n_{TV}(t) &=& \frac{1}{3} \sum_{i,k} C_{T^L_{k0} V^{(i)}_k}(t) v^{V,i}_n(t,t_0).\\
\nonumber
\tilde{C}^n_{\delta PP}(t) &=& \frac{\tilde{C}^n_{PP}(t+1)-\tilde{C}^n_{PP}(t-1)}{2a},\\
\nonumber
\tilde{C}^n_{\delta TV}(t) &=& \frac{\tilde{C}^n_{TV}(t+1) - \tilde{C}^n_{TV}(t-1)}{2a},\\
\end{eqnarray}
and their symmetric counterpart with the exchange of operators at the source and at the sink, and with the quark bilinears $P=\bar{c}\gamma_5 Q$, $A_0=\bar{c}\gamma_0\gamma_5 Q$, $V_k=\bar{c}\gamma_k Q$ and $T_{k0}=\bar{c}\gamma_k\gamma_0 Q$. In those expressions the label $L$ refers to a local interpolating field while sums over $i$ and $j$ run over the 4 Gaussian smearing levels.

\subsection{$D_s$ sector}\label{subsec-1}

\begin{figure}[t] 
\begin{center}
\begin{tabular}{cc}
	\includegraphics[width=4.5cm, clip]{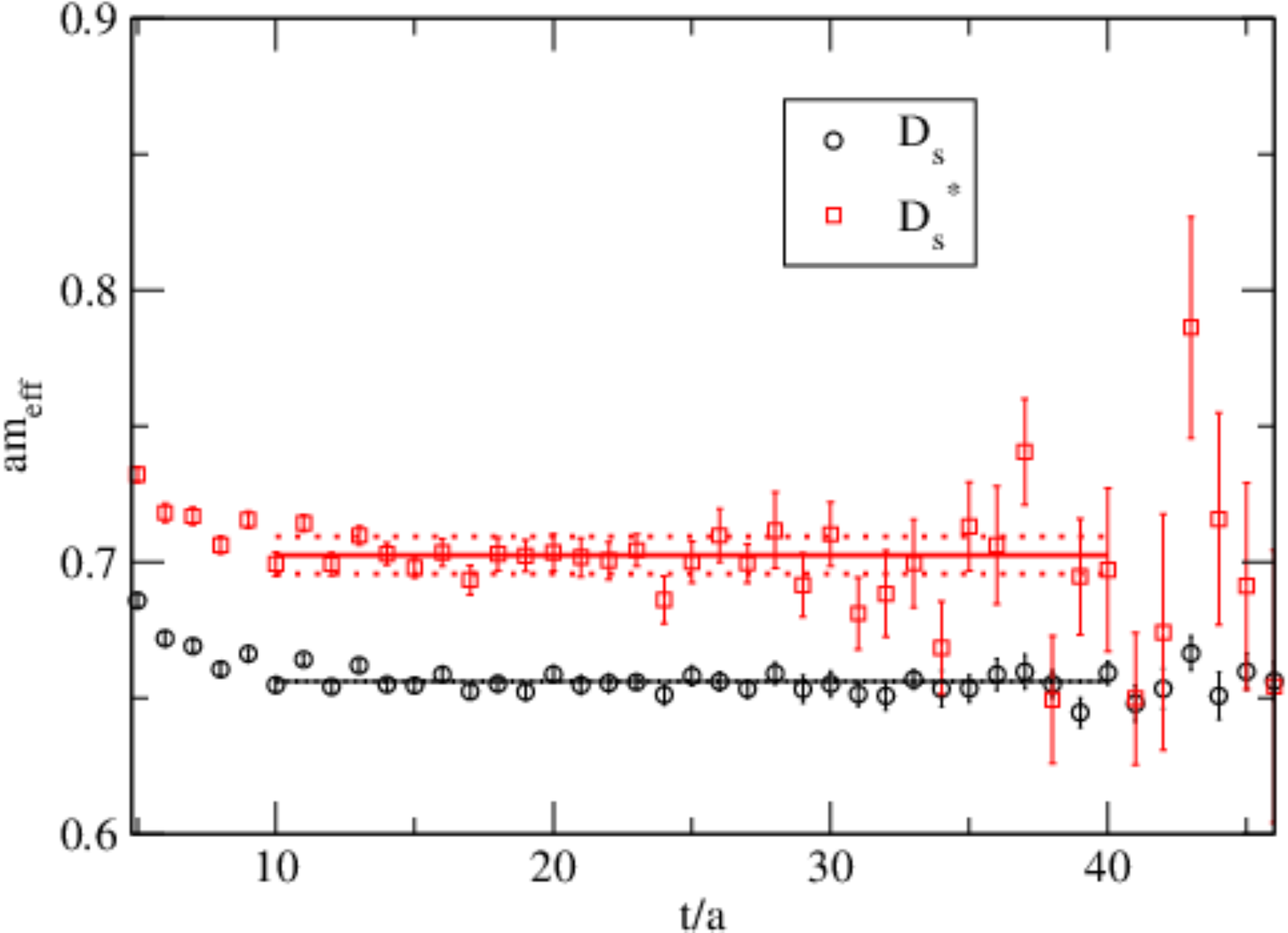}
	&
	\includegraphics[width=4.5cm, clip]{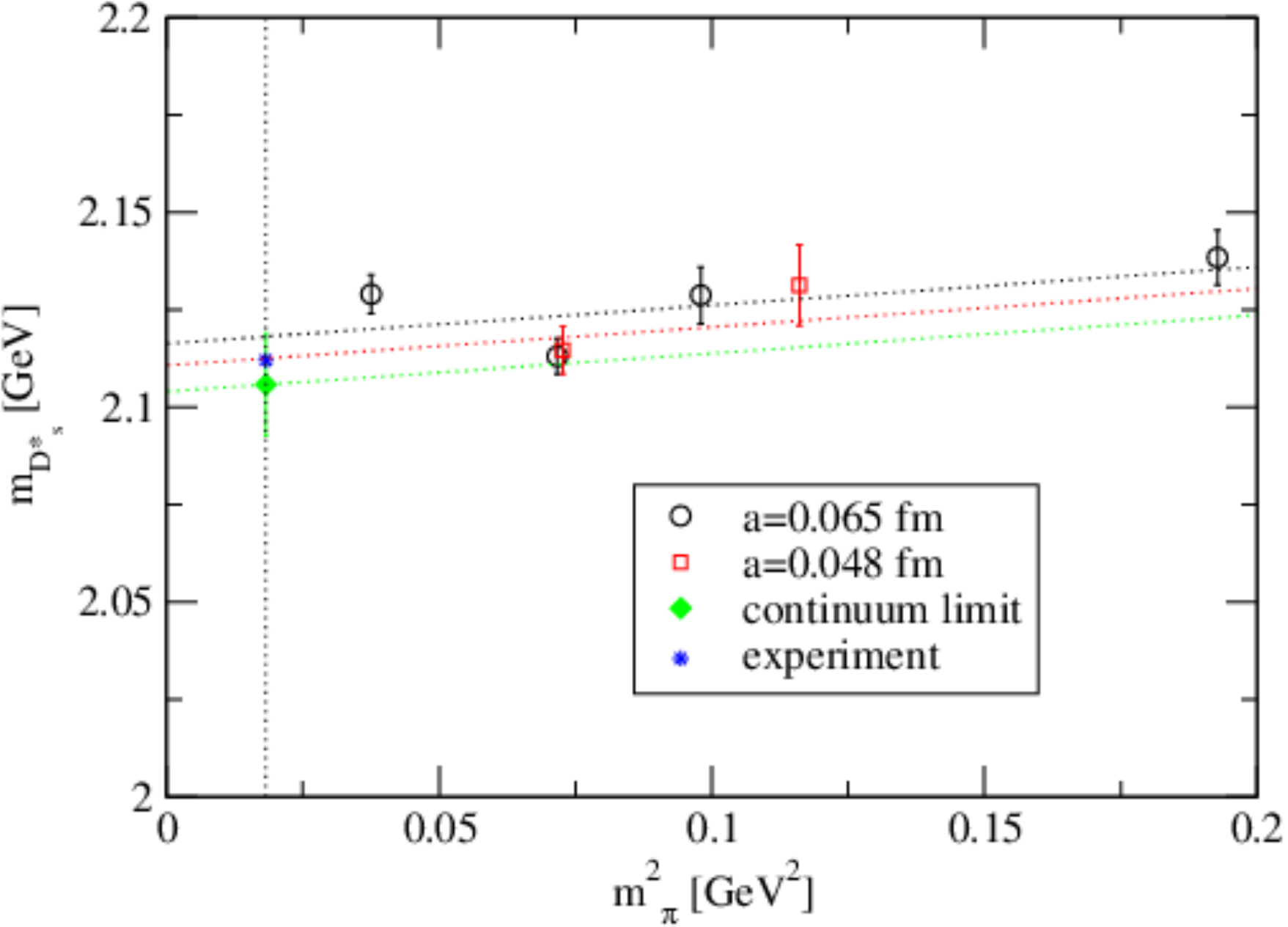}
	\end{tabular}
\end{center}
	\caption{Effective masses $am_{D_s}$ and $am_{D^*_s}$ extracted from a $4\times 4$ GEVP for the lattice ensemble F7 (left panel); we also plot the plateaus in the chosen fit interval. Extrapolation at the physical point of $m_{D^*_s}$ linear in $m^2_\pi$ and $a^2$ (right panel).}
\label{fig:massDs}
\end{figure}

\begin{figure}[t] 
\begin{center}
\begin{tabular}{ccc}
\includegraphics[width=4.5cm, clip]{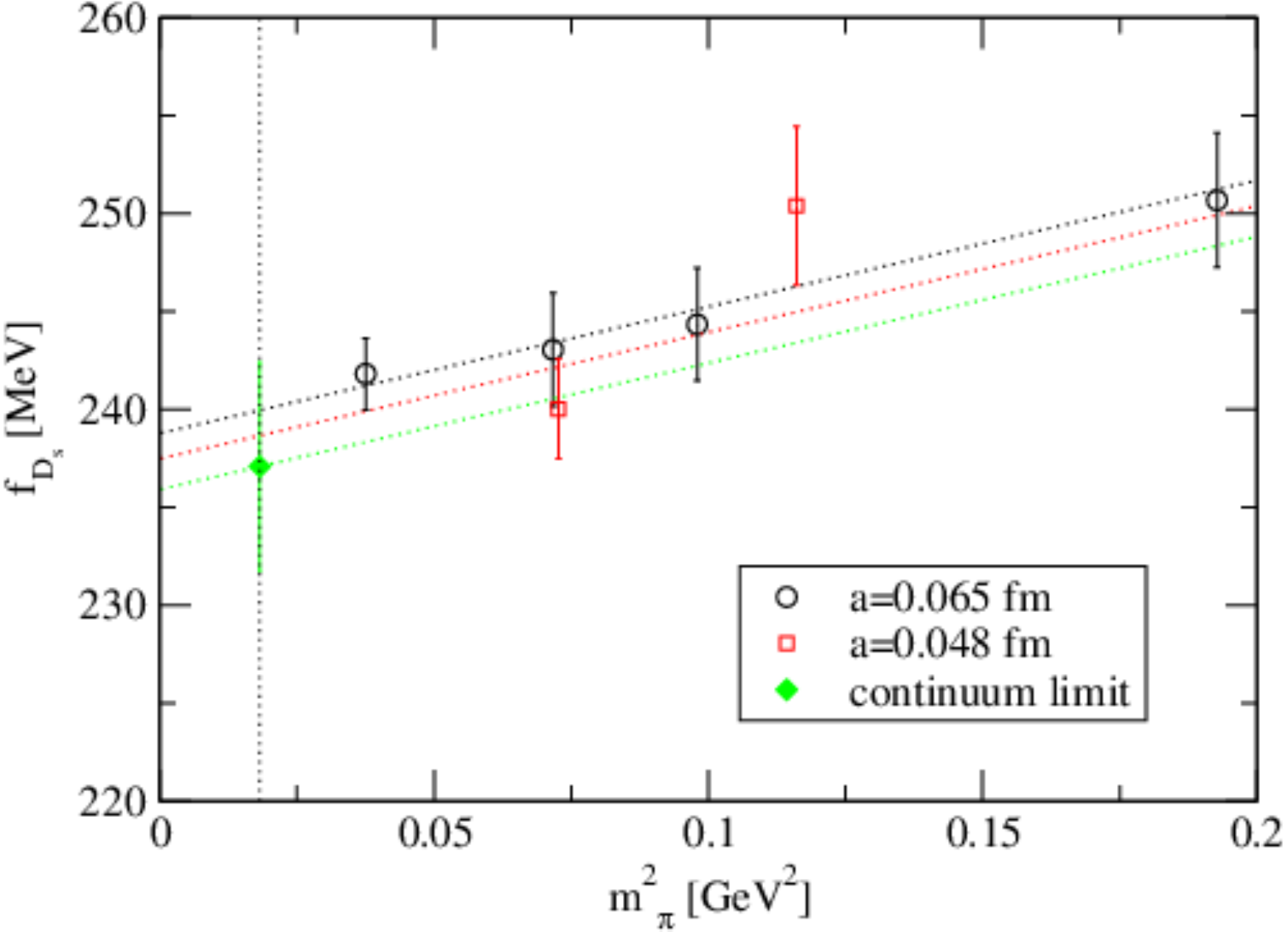}
&
\includegraphics[width=4.5cm, clip]{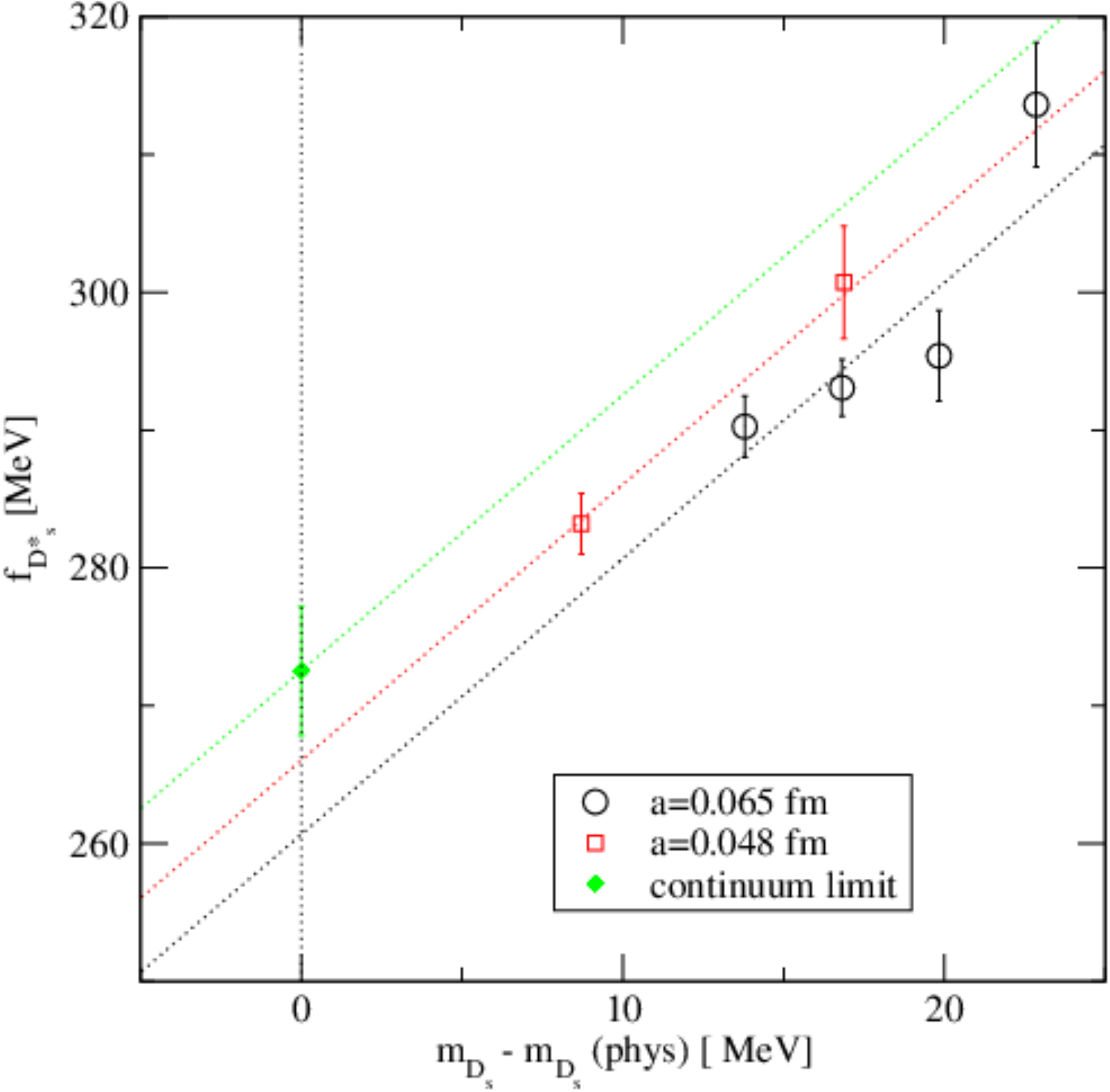}
&
\includegraphics[width=4.5cm, clip]{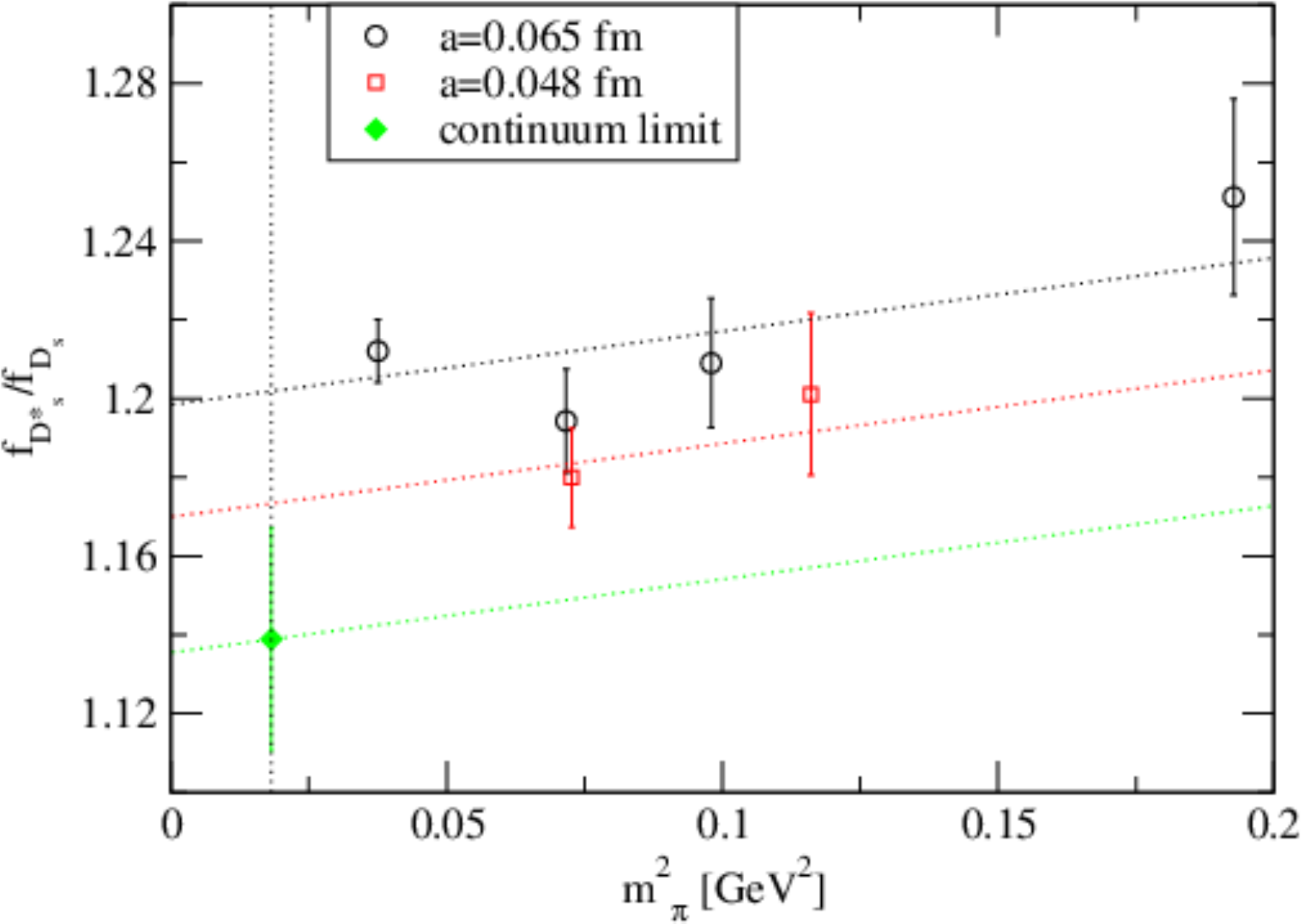}
\end{tabular}
\end{center}
	\caption{Extrapolation at the physical point of $f_{D_s}$ (left panel), $f_{D^*_s}$ (middle panel) and $f_{D^*_s}/f_{D_s}$ (right panel)by linear expressions in $m^2_\pi$ and $a^2$.}
\label{fig:decayDs}
\end{figure}

To perform the analysis of heavy-strange 2-pt correlation functions, because of large fluctuations, we have decided to use generalized eigenvectors \emph{at fixed time} $t_{\rm fix}$, $v^{P(V)}_1(t_{\rm fix}, t_0)$, to perform the corresponding projection. In practive we have chosen $t_{\rm fix}/a=t_0/a+1$ but we have checked that the results do not depend of this $t_{\rm fix}$. From the time behaviour of the projected correlators and using appropriate ratios to cancel normalization factors, we have everything to extract the matrix elements of interest, after renormalization and ${\cal O}(a)$ improvement.
In the left panel of Fig. \ref{fig:massDs} we plot the effective masses of the $D_s$ and $D^*_s$ mesons for the set F7.
%: they are obtained by the formulae
%\begin{equation*}
%am^{P,{\rm eff}}(t)={\rm argcosh} \left(\frac{\lambda^P_1(t+a,t_0)+\lambda^P_1(t-a,t_0)}{2\lambda^P_1(t,t_0)}\right),
%\quad  am^{V,{\rm eff}}(t)={\rm argcosh} \left(\frac{\lambda^V_1(t+a,t_0)+\lambda^V_1(t-a,t_0)}{2\lambda^V_1(t,t_0)}\right).
%\end{equation*}
One can see that our plateaus are satisfying. As shown in the right panel of Fig. \ref{fig:massDs} we have checked that at the physical point $m_{D^*_s}$ is compatible with the experimental value 2.112 GeV, with cut-off effects limited to 0.5\% at $\beta=5.3$; we have obtained 
$m_{D^*_s}=2.106(13)(13)%(8)
\,{\rm GeV}$,
where the first error is statistical and the second error accounts for the uncertainty on the lattice spacing. 
%and the third error counrs for a next-to-leading order term in the chiral fit, whose the coefficient is however compatible with 0. 
Extrapolations at the physical point of $f_{D_s}$, $f_{D^*_s}$ and $f_{D^*_s}/f_{D_s}$ are displayed in Fig. \ref{fig:decayDs}: done linearly in $m^2_\pi$ and $a^2$, they are all quite mild. Cut-off effects on $f_{D_s}$ are limited to 1\% at $\beta=5.3$ while they are quite stronger for $f_{D^*_s}$, of the order of 7\%: they propagate in the ratio $f_{D^*_s}/f_{D_s}$ with an effect of 6\%. We will quote as the main preliminary result the ratio
\begin{equation}\nonumber
f_{D^*_s}/f_{D_s}=1.14(3),
%(1),
\end{equation}
where the systematic error coming from the uncertainty on lattice spacings is negligible.
%; again a next-to-leading order contribution
%in the chiral fit has a coefficient compatible with 0 but we included conservatively the discrepancy on results at the physical point in the 
%systematic error.

\begin{figure}[t] 
\begin{center}	
\includegraphics*[width=7cm,clip]{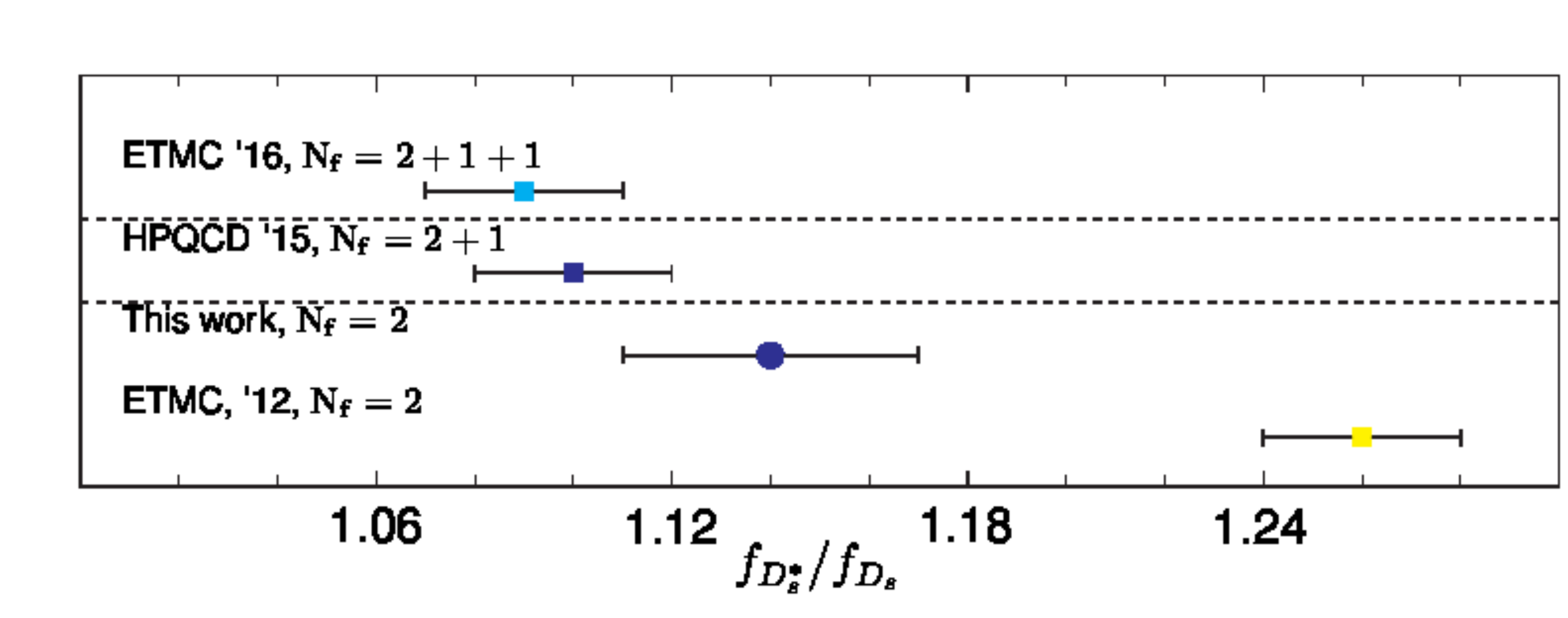}
\end{center}
	\caption{Collection of lattice results of $f_{D^*_s}/f_{D_s}$.}
\label{fig:collectionfDs}
\end{figure}

\noindent
So far there are only 2 lattice estimates of $f_{D^*_s}/f_{D_s}$ at $N_f=2$ by ETMC \cite{Becirevic:2012ti} and us, and 2 other have been performed at $N_f=2+1$ by HPQCD \cite{Donald:2013sra} and $N_f=2+1+1$ by ETMC \cite{Lubicz:2017asp}. We collect the various results in Fig. \ref{fig:collectionfDs}. In the past it was thought it could be a quantity where quite large quenching effects of the strange quark show up with an amount larger than 10\%, because the first result from ETMC is around 1.25. Our finding would tend to the conclusion that it is less pronounced: still, the trend is to have less spin breaking effects when more flavours are active.

\subsection{Charmonia sector}\label{subsec-2}

\begin{figure}[t] 
\begin{center}
\begin{tabular}{cc}
	\includegraphics[width=4cm,clip]{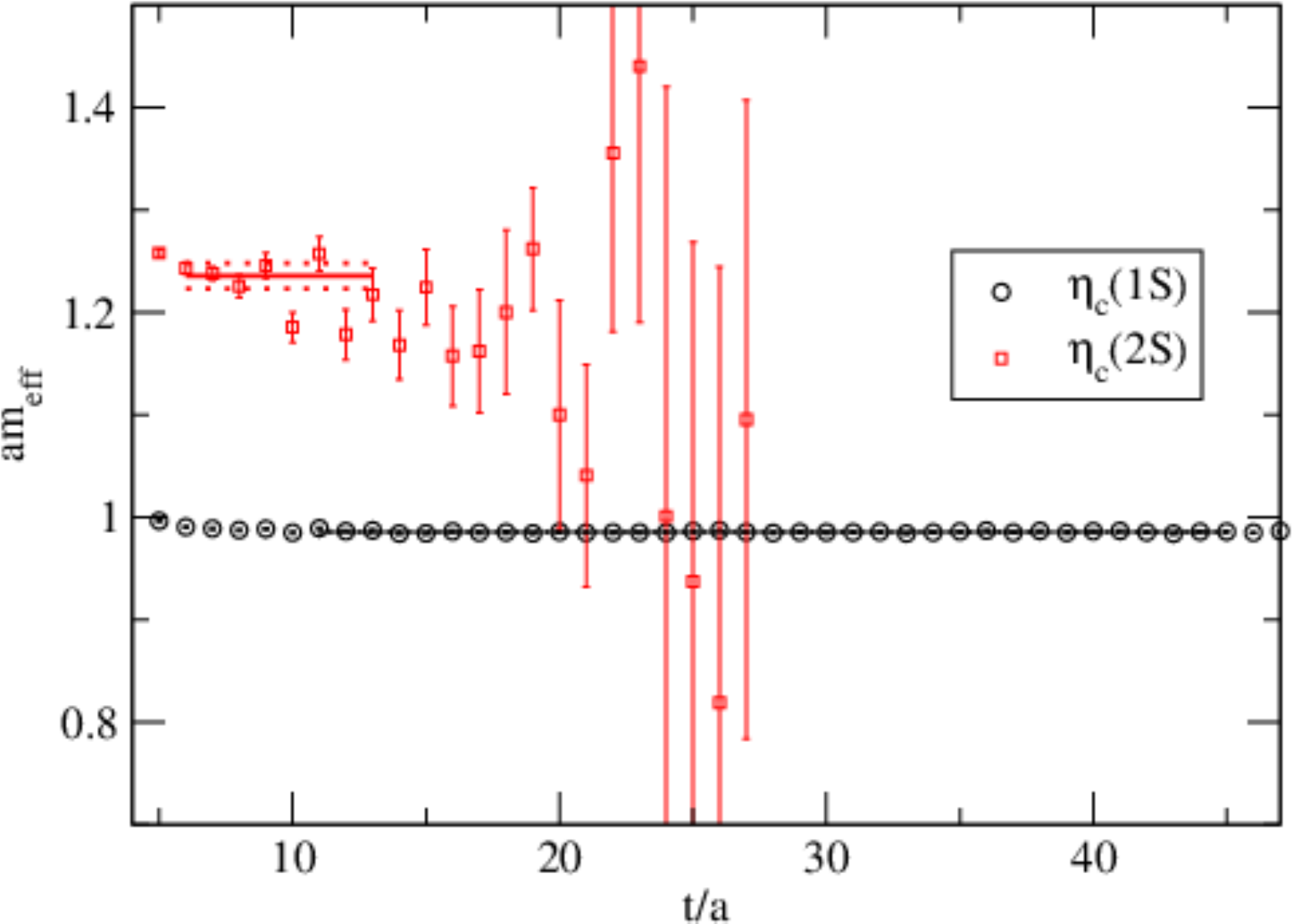}
&
	\includegraphics[width=4cm]{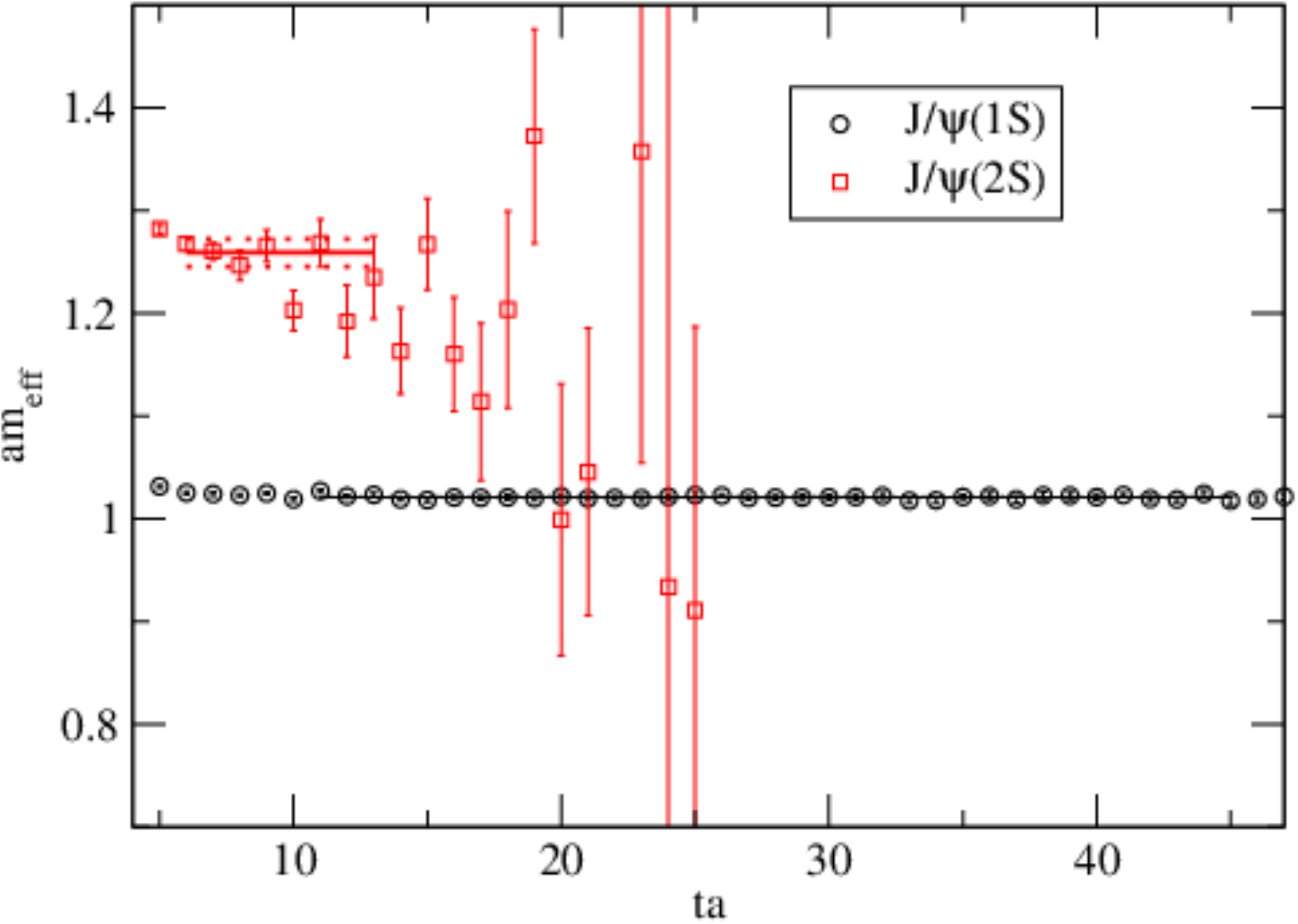}
\end{tabular}
\end{center}
	\caption{Effective masses $am_{\eta_c}$ and $am_{\eta_c(2S)}$ (left panel), $am_{J/\psi}$ and $am_{\psi(2S)}$ (right panel) extracted from a $4\times 4$ GEVP for the lattice ensemble F7; we also plot the plateaus in the chosen fit interval.}
\label{fig:massescharmonia}
\end{figure}

\begin{figure}[t] 
\begin{center}
\begin{tabular}{cc}
\includegraphics[width=5cm, clip]{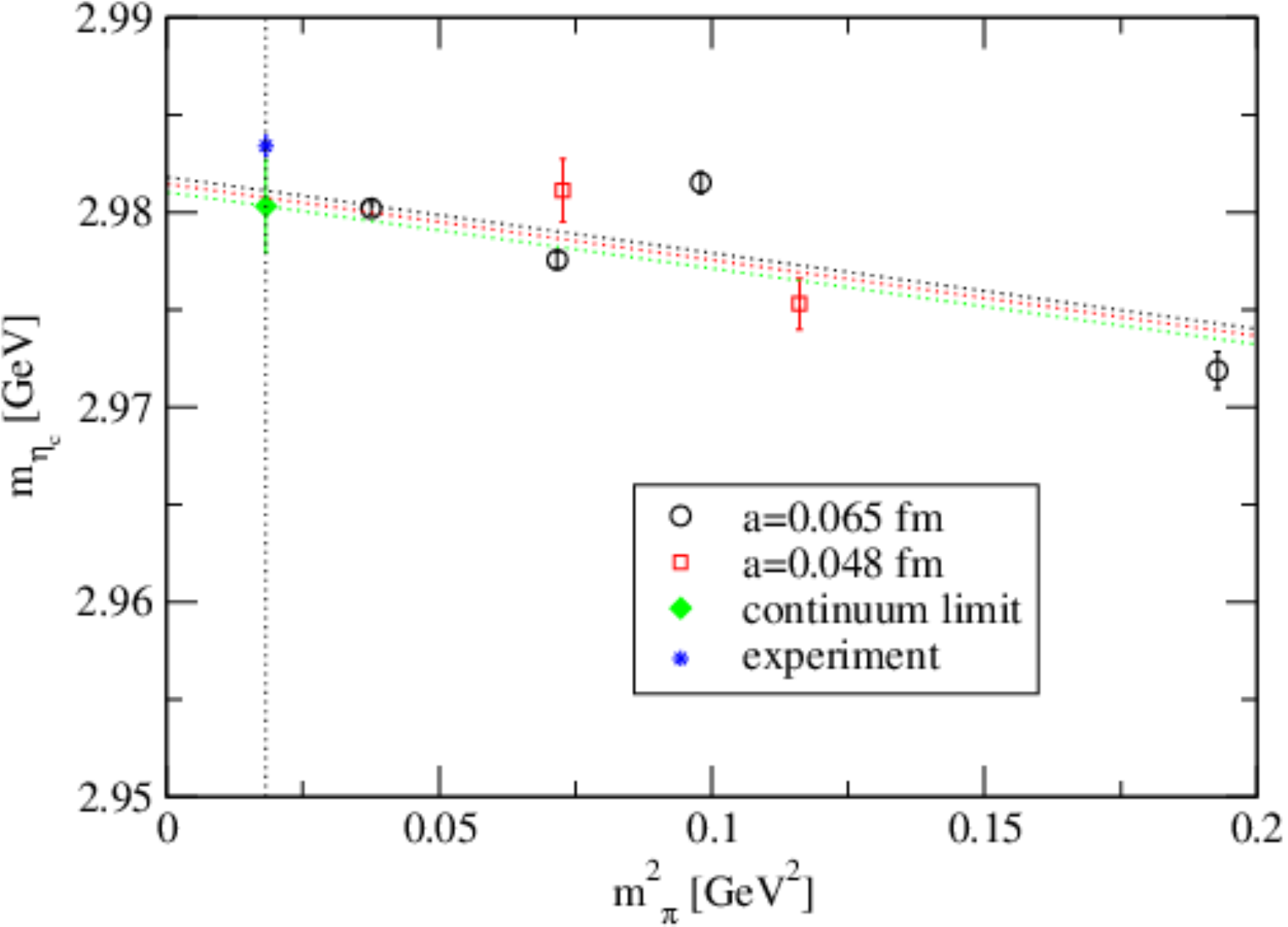}
&
\includegraphics[width=5cm, clip]{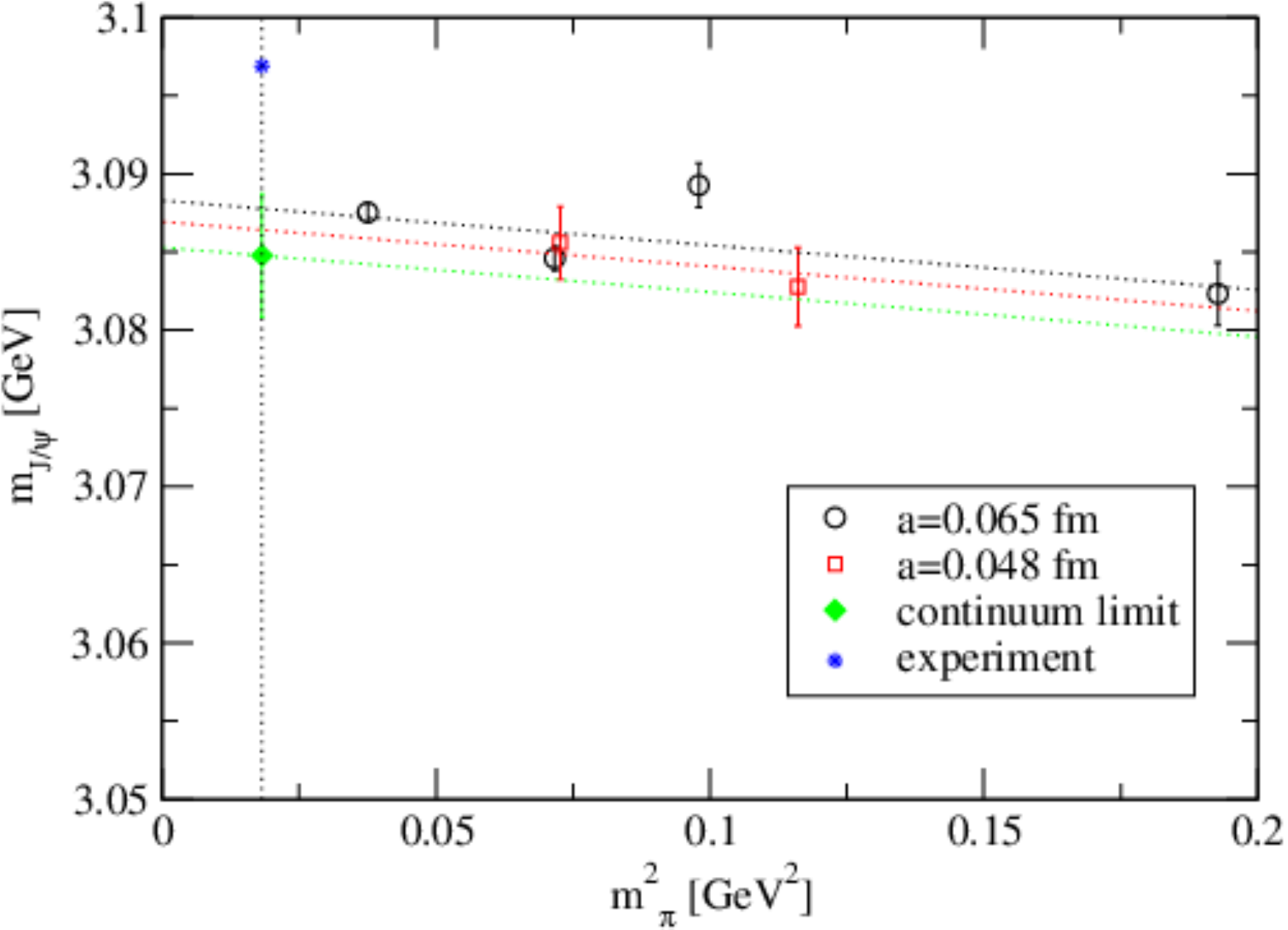}
\end{tabular}
\end{center}
	\caption{Extrapolation at the physical point of $m_{\eta_c}$ (left panel) and $m_{J/\psi}$ (right panel) by linear expressions in $m^2_\pi$ and $a^2$.}
\label{fig:massphysical}
\end{figure}

\begin{figure}[t] 
\begin{center}
\begin{tabular}{cc}
\includegraphics[width=5cm, clip]{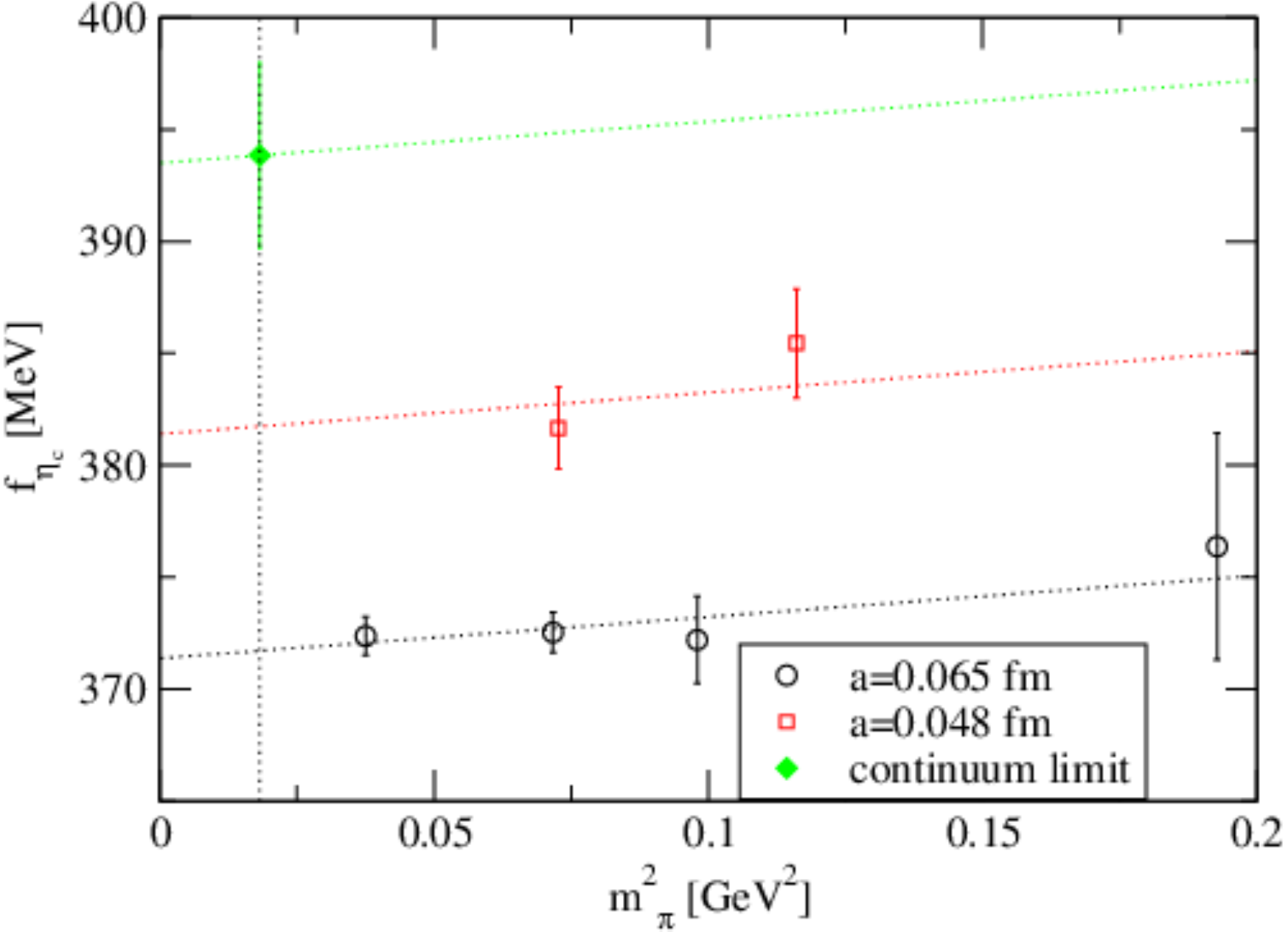}
&
\includegraphics[width=5cm, clip]{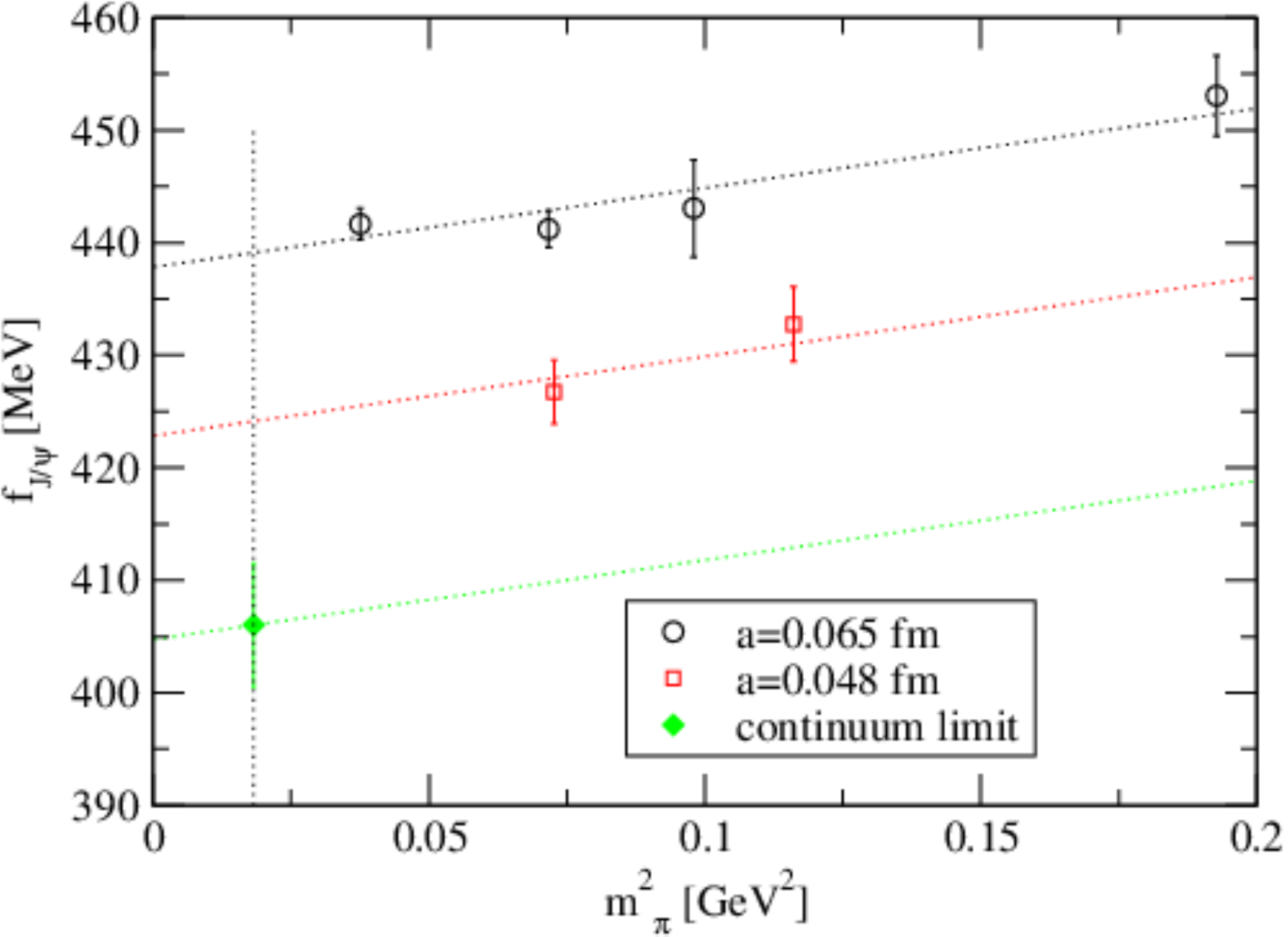}
\end{tabular}
\end{center}
	\caption{Extrapolation at the physical point of $f_{\eta_c}$ (left panel) and $f_{J/\psi}$ (right panel) by linear expressions in $m^2_\pi$ and $a^2$.}
\label{fig:decay}
\end{figure}

We proceed in the same way as in the previous subsection to extract proterties of pseudoscalar and vector charmonia, including those of
radial excitations. In Fig. \ref{fig:massescharmonia} we plot effective masses of the $\eta_c$, $\eta_c(2S)$, $J/\psi$ and $\psi(2S)$ mesons for the set F7.
%: they are obtained by the expressions
%\begin{equation*}
%am_{\eta_c}^{\rm eff}(t)={\rm argcosh} \left(\frac{\lambda^P_1(t+a,t_0)+\lambda^P_1(t-a,t_0)}{2\lambda^P_1(t,t_0)}\right),
%\quad  am_{J/\psi}^{\rm eff}(t)={\rm argcosh} \left(\frac{\lambda^V_1(t+a,t_0)+\lambda^V_1(t-a,t_0)}{2\lambda^V_1(t,t_0)}\right).
%\end{equation*}
%\begin{equation*}
%am_{\eta_c(2S)}^{\rm eff}(t)={\rm argcosh} \left(\frac{\lambda^P_2(t+a,t_0)+\lambda^P_2(t-a,t_0)}{2\lambda^P_1(t,t_0)}\right),
%\quad  am_{\psi(2S)}^{\rm eff}(t)={\rm argcosh} \left(\frac{\lambda^V_2(t+a,t_0)+\lambda^V_2(t-a,t_0)}{2\lambda^V_1(t,t_0)}\right).
%\end{equation*}
One can see that our plateaus are pretty long for the ground states but are unfortunately shorter for the radial excitations. The latter are acceptable
for a qualitative exploration but not for a precision measurement.
We show in Fig. \ref{fig:massphysical} the extrapolation to the physical point of $m_{\eta_c}$ and $m_{J/\psi}$: the dependence on $m^2_\pi$ and $a^2$ is mild, with cut-off effects almost negligible. However the contribution to the meson masses besides the mass term $2m_c$ is difficult to catch. At the physical point $m_{\eta_c}$ and $m_{J/\psi}$ are compatible with the experimental values 2.983 GeV and 3.097 GeV: 
$m_{\eta_c}=2.980(2)(18)$ GeV and $m_{J/\psi}=3.085(4)(19)$ GeV,
%\begin{equation*}
%m_{\eta_c}=2.982(1)(19)\,{\rm GeV}, \quad m_{J/\psi}=3.084(4)(19)\, {\rm GeV},
%\end{equation*}
where the first error is statistical and the second error accounts for the uncertainty on the lattice spacing: the latter clearly dominates and hides a possible mismatch between our extrapolated results at the physical point and experiment. We display in Fig. \ref{fig:decay} extrapolations at the physical point of $f_{\eta_c}$ and $f_{J/\psi}$: they are mild, cut-off effects on $f_{\eta_c}$ are of the order of 4\% at $\beta=5.3$ while they are stronger for $f_{J/\psi}$, about 10\%. We get as preliminary results
\begin{equation}\nonumber
f_{\eta_c}=394(4)(2)\, {\rm MeV}, \quad f_{J/\psi}=406(5)(3)\, {\rm MeV},
%f_{\eta_c}=386(3)(2)\, {\rm MeV}, \quad f_{J/\psi}=399(4)(3)\, {\rm MeV},
\end{equation}
where the systematic error comes from the uncertainty on lattice spacings.
One can derive a phenomenological estimate of $f_{J/\psi}$. Indeed, using the expression of the electronic decay width
\begin{equation}\nonumber
\Gamma(J/\psi \to e^+e^-)=\frac{4\pi}{3}\frac{4}{9} \alpha(m^2_c)\frac{f^2_{J/\psi}}{m^2_{J/\psi}},
\end{equation}
the experimental determination of the $J/\psi$ mass and width and setting $\alpha_{\rm em}(m^2_c)=\frac{1}{134}$, one gets
$f^{``{\rm exp}"}=407(6)$ MeV.

\begin{figure}[t] 
\begin{center}
\begin{tabular}{cc}
	\includegraphics[width=7cm,clip]{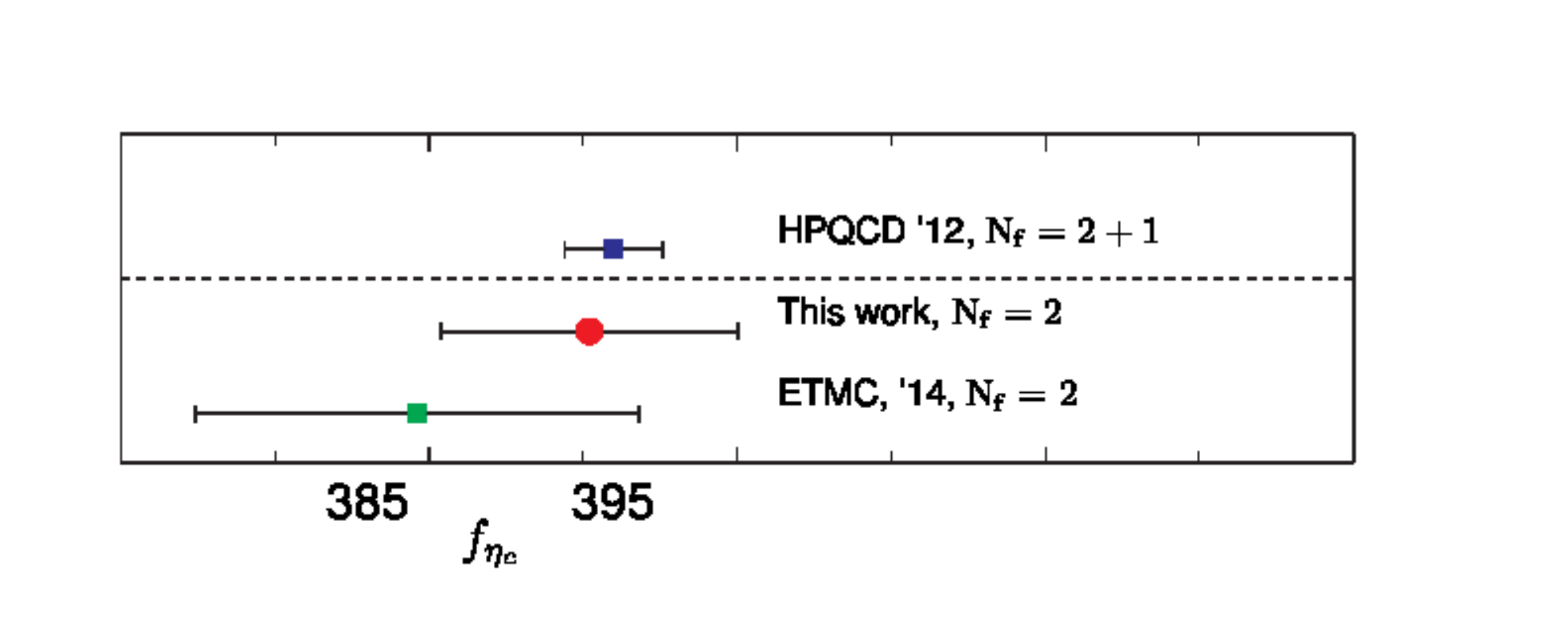}
&
	\includegraphics[width=7cm,clip]{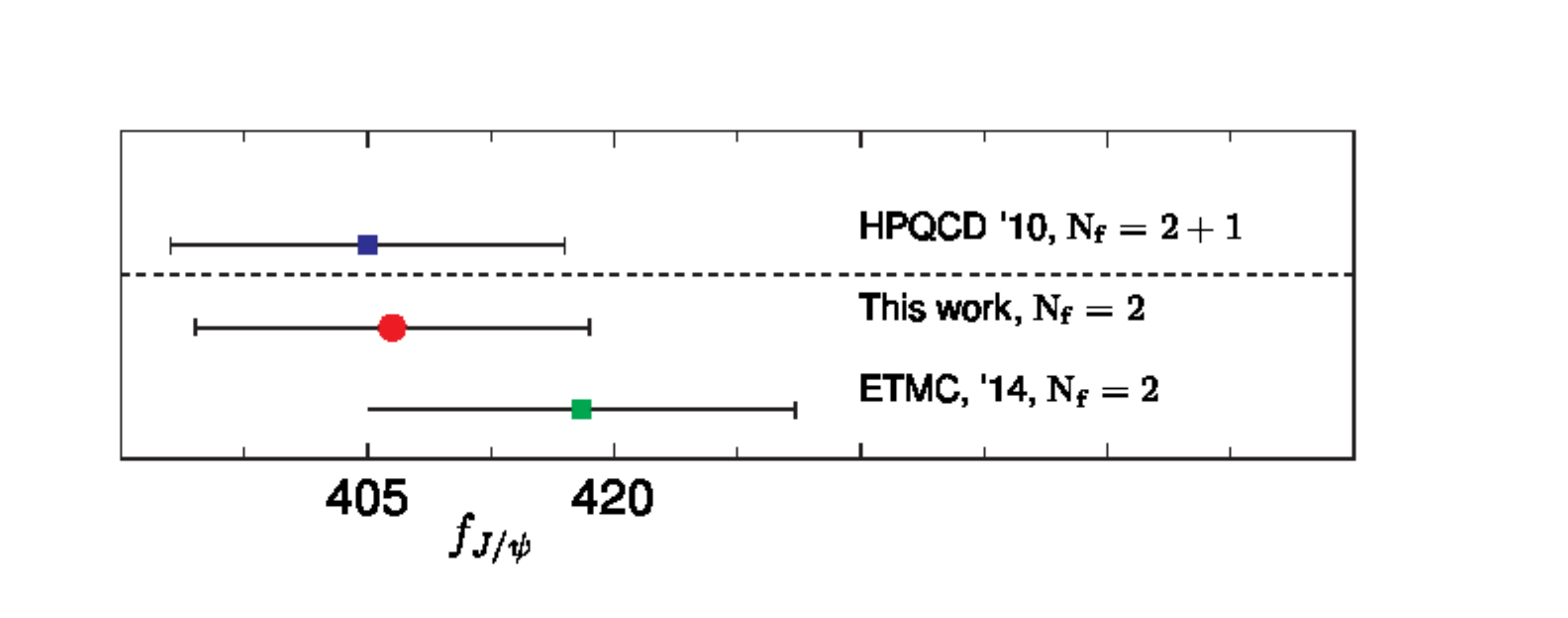}
\end{tabular}
\end{center}
	\caption{Collection of lattice results of $f_{\eta_c}$ (left panel) and $f_{J/\psi}$ (right panel).}
\label{fig:collectfetac}
\end{figure}

\noindent 
So far there are only 2 lattice estimates of the $\eta_c$ and $J/\psi$ charmonia decay constants at $N_f=2$ by ETMC \cite{Becirevic:2013bsa} and us, and  a third computation have been performed at $N_f=2+1$ by HPQCD \cite{Colquhoun:2014ica}. We collect the various results in Fig. \ref{fig:collectfetac}.

\noindent
Unfortunately the situation is not as promising for radial excited states. With small cut-off effects, of the order of 5\%, we have obtained
%In Fig. \ref{fig:ratiomass} we show the extrapolation to the continuum limit
%of the ratios $m_{\eta_c(2S)}/m_{\eta_c}$ and $m_{\psi(2S)}/m_{J/\psi}$, compared to the experimental values. As the cut-off effects
%are small, of the order of  5\%, there is no hope to have points in the continuum limit significantly lower than our lattice data: we get 
$m_{\eta_c(2S)}/m_{\eta_c} \gg (m_{\eta_c(2S)}/m_{\eta_c})^{\rm exp}$ and $m_{\psi(2S)}/m_{J/\psi} \gg (m_{\psi(2S)}/m_{J/\psi})^{\rm exp}$.
%In \cite{damircharm} lattice results were much closer to the experiment but the slope in $a^2$ was probably overrestimated from the coarsest 
%lattice point. 
The situation is even more confusing for the ratios of decay constants
%, as shown in Fig. \ref{fig:ratiodecay}
: $f_{\eta_c(2S)}/f_{\eta_c} < 1$ while $f_{\psi(2S)}/f_{J/\psi} >1$.

\section*{Acknowledgments} This work was granted access to the HPC resources of CINES and IDRIS under the allocations 2016-x2016056808 
and 2017-A0010506808 made by GENCI. This work was partly supported by the grant {HE~4517/3-1}
(J.~H. and M.~P.) of the Deutsche Forschungsgemeinschaft.

%\clearpage
\bibliography{Lattice2017_114_BLOSSIER}

%%%%%%%%%%%%%%%%%%%%%%%%%%%%%%%%%%%%%%%%%%%%%%%%%%%%%%%%%%%%%%%%%%%%%%%%%%%%%
\end{document}